\documentclass{PoS}

\title{X-ray Radio Correlation In  Black Hole Sources}

\ShortTitle{X-ray Radio Correlation In  Black Hole Sources}

\author{\speaker{A. R. Rao}
         \\
        TATA Institute of Fundamental Research, Mumbai, India\\
        E-mail: \email{arrao@tifr.res.in}}

\abstract{  
We examine the X-ray - radio correlation in Galactic black hole sources. 
We highlight some of the results which extend the flux-flux relations  to
 sources with very high accretion rates. Some of the recent results indicate that
 the synchrotron process is unlikely to be the mechanism responsible for the X-ray emission,
 particularly at high accretion rates. We present a truncated accretion disk scenario
 and argue that accretion rate and accretion disk geometry ultimately act as a
 driver of the X-ray - radio correlation. We stress the importance of wide-band
 X-ray spectral measurements to understand the disk-jet connection and briefly outline
 some attempts made in the Indian context to build instruments for wide-band X-ray spectroscopy.}
\FullConference{VI Microquasar Workshop: Microquasars and Beyond\\
		 September 18-22, 2006\\
		 Como, Italy}

\begin{document}
\section{Introduction}

Ever since the discovery of the superluminally moving radio ejecta from the Galactic black hole 
source GRS 1915+105 \cite{mirrod94}, there is a growing realization of a close
 connection between the underlying Physics of accretion in Active Galactic Nuclei (AGN)
 and Galactic black hole sources, and, these days, the name `microquasar`
 is not primarily restricted to sources  with measurable ejecta, but rather encompasses 
the whole class of Galactic black hole sources. Traditionally, superluminal jets are
 studied in the context of  AGNs and probably due to the lower characteristic accretion
  disk temperature in these sources (a few eV) and the consequent atomic physics interplay
 in the surrounding medium, study of AGNs are primarily centered around the investigation
 of the AGN environment. On the other hand, study of Galactic black hole sources are synonymous
 with the study of some of the brightest X-ray sources in the sky and the X-ray emission
  is directly related to the innermost parts of the accretion disk. Hence the field of
 micro-quasars is an important bridge between these two approaches and has provided several
 insights to the `disk-jet` connection which probably would lead to a good understanding of
 the Physics in the extreme conditions near the event horizon of a black hole.

One of the important findings of the past decade of research is the ubiquity of jet emission.
 All classes of accreting sources
 are observed to be having a measurable radio emission, presumably arising from the putative
 jet \cite{fenkuu01}. Further, there was a broad trend of association of radio
 emission with the X-ray emission, which prompted several workers to look for a fundamental
 plane of black hole activity involving the X-ray luminosity (L$_X$), radio luminosity(L$_R$)
 and the black hole mass (M$_{BH}$) and a relation between them is obtained as
\cite{merl03, falc04}:
$$Log (L_R) = 0.6 Log (L_X)  + 0.78 Log (M_{BH})$$
Though this relation was found to be valid for fifteen orders of magnitude variation
  in luminosities (from Galactic black hole sources to distant quasars), it was pointed out
{}   that most of this correlation is due to the distance effect \cite{breg05}. If one examines
 the correlation in the flux plane \cite{merl06}, it is seen that a correlation
 does exist among these parameters (with a reduced range of variation - 3 to 4 orders of
 magnitude change in the flux values), but the scatter in the correlation indicating
 a broad trend of variation rather than any fundamental plane of activity.

Here we examine the X-ray radio correlation in Galactic black hole sources with a aim
 to understand the accretion  phenomena that is responsible for the disk structure as well as
 the jet formation. We briefly examine the flux-flux relations and highlight
 some of the efforts to extend this relation to sources with very
 high accretion rates. From the nature of the wide band X-ray spectrum
 at these high accretion rates, we conclude that synchrotron emission is unlikely
to be the radiation process responsible for the X-ray emission,
 particularly at high accretion rates. Formation of a truncated accretion disk
 appears to be responsible for the jet emission. Accretion rate and accretion disk
 geometry ultimately decides the nature of the truncated accretion disk resulting in an
 X-ray-radio correlation in these sources. We outline a scenario for the superluminal
 jet emission  and finally stress the importance of wide-band X-ray spectral measurements
 to understand the disk-jet connection. Some attempts made in the Indian
 context to build instruments for wide-band X-ray spectroscopy are briefly outlined.

\section{Flux - flux correlations}

A broad trend of increasing radio emission with a corresponding increase in  X-ray
 emission was noticed in the Galactic black hole source GX 339-4
   \cite{hann98}. The source was monitored at 843 MHz
  for about three years (1994-97) and during this period monitoring hard X-ray
 data were available from BATSE onboard CGRO.  The radio monitoring was more extensive after
 1995 and during this period soft X-ray monitoring data were also available from ASM onboard
 the RXTE satellite. A very strong positive correlation was noticed between radio emission and
 the  soft X-ray emission as well as between radio emission and the hard X-ray emission.

Such correlations were extended in this source and several other sources \cite{broc99, corb01}.
Fender \cite{fend01} in fact concluded that jet emission in the low-hard states
 of Galactic black hole sources constitute a substantial fraction  (several percent)
of the accretion energy. Gallo et al. \cite{gall03}
 collated all such observations and demonstrated a   strong correlation   between
 the soft X-ray luminosity and radio luminosity for nine black hole sources ranging
 all the way from the `off` state to almost upto the transition to `high-soft` state
 (from 10$^{-6}$ L$_{Edd}$ to 0.02 L$_{Edd}$). A quenching of the radio emission was also noticed in
 the `high-soft` state of some of the sources.

This correlation was extended to even higher luminosities by Choudhury et al. \cite{chou03}
  by including the bright sources Cygnus X-3 and GRS 1915+105. Choudhury et al \cite{chou02}
 found a strong correlation  between the soft X-ray emission and the radio emission
 during the relatively steady state of Cyg X-3 (which they called the `low-hard` state)
 and also a strong anti-correlation between the radio emission and the hard X-ray emission.
 This study was extended to include the bright micro-quasar GRS 1915+105, and, it was found that
 the relation obtained by Gallo et al. \cite{gall03} could be extended almost upto the Eddington
 luminosity (see Figure 1). One of the interesting findings of Choudhury et al. \cite{chou03}
 is that the radio X-ray correlations are likely to be due to the spectral pivoting
 seen in Cyg X-3 and GRS 1915+105.  

\begin{figure}
\begin{center}
\includegraphics[angle=0,width=.8\textwidth]{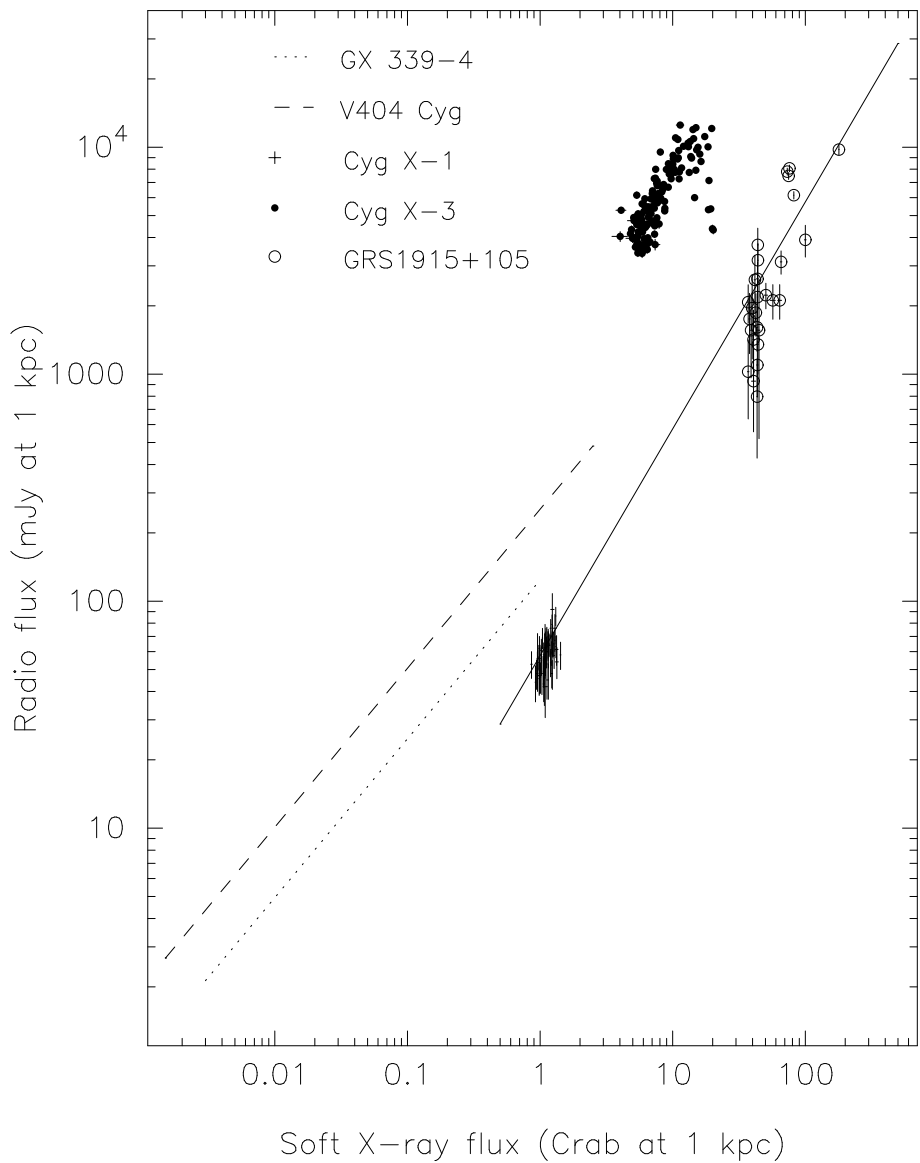}
\caption{
X-ray - radio correlation in Galactic black hole sources. Figure adapted
from Choudhury et al. \cite{chou03}. 
}
\label{xradio}
\end{center}
\end{figure}

\begin{figure}
\begin{center}
\includegraphics[angle=270,width=.8\textwidth]{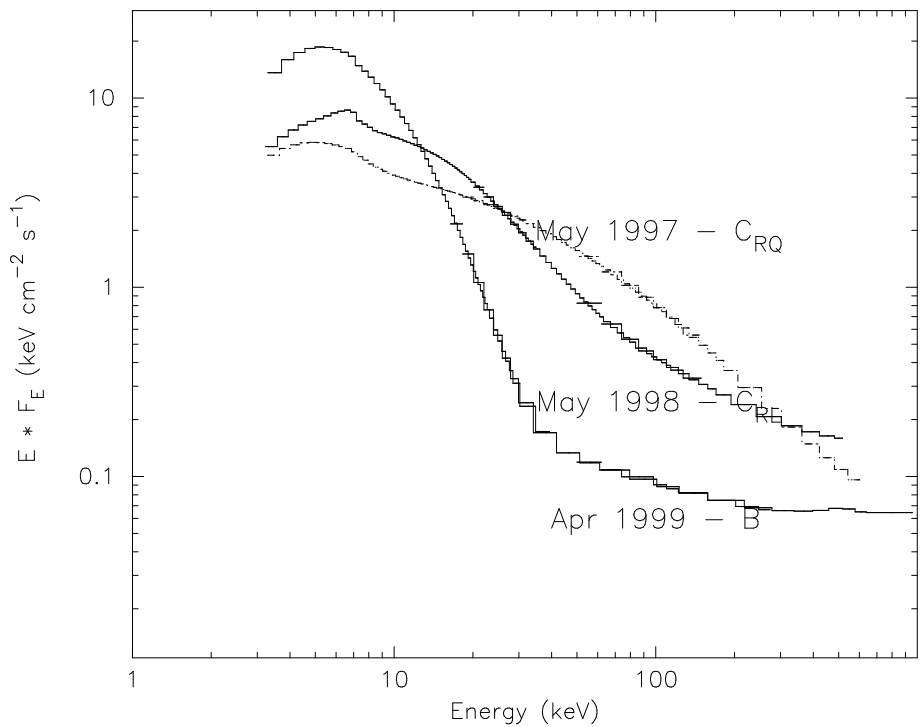}
\caption{
Wide band X-ray spectra of GRS 1915+105 in high soft state (state B - Apr 1999),
radio-loud steady state (state C$_{RL}$ - May 1998) and radio-quiet steady state (state C$_{RQ}$ - May 1997).
}
\label{grswide}
\end{center}
\end{figure}

\section{Synchrotron radiation as the driver of the X-ray - radio  correlation ?}

The X-ray radio correlation seen in the `low-hard` states of the Galactic black hole
 sources had prompted several workers to investigate whether both the radiations are
 produced by the same physical process. Since the inverted spectral shape and 
 the size of the radio emission showed that the radio emission is due to synchrotron
 radiation, the possibility of the X-ray emission originating from the same physical process was
 examined in detail. Vadawale et al. \cite{vada01} investigated whether the power-law spectral
 component seen in GRS 1915+105 could be due to synchrotron radiation from the base of the jet.
 Though they could fit the observed radio emission using a simplified model, they found that
 the inverted radio spectrum is too steep due to assumption of  adiabatic  expansion  and the
 total jet power exceeds accretion power.  Much more specific models have been developed
\cite{mark01},
and in some cases the synchrotron models are claimed to be
 successful in explaining the radio-X-ray spectrum at low accretion rates 
\cite{mark05}.

There are, however, several evidences which indicate that the synchrotron radiation
 is unlikely to be responsible for the X-ray emission, particularly at high accretion rates.
 Wide band spectral fit for 1E 1740.7-2942  using a synchrotron model required extreme parameters
 to explain the X-ray emission and still over-predicts the radio emission \cite{bosc06}. Examining the
 wide band X-ray/ gamma-ray spectra of several black hole sources Zdziarski et al. 
\cite{zdzi03} 
 concluded that the relation between level of X-ray emission and rate of jet production is
 related to a common accretion mechanism. During the hard state of XTE J1550-564
Yuan et al. \cite{yuan06}
detected a secondary maximum in optical (which is due to
 synchrotron radiation), but not seen in X rays, implying that the X-ray emission from
  jets is not an  important contributor to the observed X-ray spectrum.
 The spectral energy distribution (SED) of XTE J1550-564  was examined in detail by
 extrapolating the physical parameters derived from the direct imaging of
soft X-ray emitting blobs and it was concluded the  synchrotron mechanism
 contributes little to the observed SED at the X-ray region, particularly  when the
 source is relatively bright \cite{xue06}. Hence we can conclude that synchrotron process is unlikely
 to contribute towards the X-ray emission of black hole sources and hence is not the
 driving mechanism for the X-ray-radio correlation seen in these sources.  

\section{Truncated Accretion Disc}

There is a growing evidence that a truncated accretion disk exists in black hole sources,
 particularly in the intermediate state (or the very high state).
  The wide band X-ray spectrum of GRS 1915+105 (see Figure 2) in the steady
 state shows three spectral components: a thermal component (modeled as a disk black
 body emission), a highly saturated Comptonized component and a power law. A
 detailed spectral analysis using quasi-simultaneous low energy gamma-ray data from
 the OSSE detector onboard the CGRO 
\cite{zdzi01, zdzi05}
identified the hard spectral components as due to Comptonization of a plasma containing electrons
 of thermal as well as non-thermal energy distribution. 

The existence of the two hard spectral components in the very high state of GRS 1915+105
 was made the basis for a qualitative concept to explain the radio X-ray correlation in
 Galactic black hole sources \cite{chou03}.
Briefly, during the low-hard state
 of a source, the accretion disk truncates at a large distance from the black hole and
 the material within this region is filled with the low density high temperature plasma
 (kT$_e$ $\sim$ 50 - 100 keV; $\tau$ < 1, as usually seen in such states). The amount of jet emission
 depends on the radius of the truncated accretion disk. As the accretion rate increases,
 the truncation radius decreases, giving rise to enhanced emissions in the  X-ray and radio
 wavebands. During high accretion rates the truncation radius comes very close to the
 black hole and the plasma becomes cool and dense. Non-thermal phenomena too start taking
 place. In such high accretion rates, the radio as well as soft X-ray emissions are high,
 but there would be pivoting at around a few tens of keV giving rise to an anticorrelation
 between soft and hard X rays. When the accretion rate increases beyond a critical limit,
 the accretion disk would reach upto the last stable orbit without having any
 truncation radius, thus quenching the radio emission.

The change in luminosities will occur in the viscous time scale corresponding to
 the truncation radius - a couple of thousand seconds for about 10 Schwartzschild
 radius  (R$_S$)  in a  ten solar mass black hole \cite{bell97}.
  A change in the accretion phenomena will result in a corresponding change in the
 Comptonizing plasma in Compton cooling time scale or electron heating time scale
 which is of the order of a few hundred seconds
\cite{chaman00}.
 One should see a corresponding change in the radio luminosity in the jet production time
 scale, which is also about a few hundred second to a few thousand seconds
\cite{mira98}.
Hence, in the scenario of a truncated accretion disk
 driving the X-ray radio correlation the following time scales should manifest
 themselves, particularly in the very high state of the source: the truncation disk
 changes its size in about a couple of thousand
seconds giving rise to an increased soft X-ray emission (if the truncation
 radius decreases); there would be a corresponding increase in the radio emission from
 the jet, after a delay of the jet production time scale; thermal plasma would be
 Compton cooled due to the enhanced number of seed photons,  giving rise to a
 decrease in hard X-ray emission with a positive delay.

Choudhury \& Rao \cite{chorao04}  detected a delay between the soft and hard X-ray emission
 in    Cygnus X-3.  Similar delays were also found in GRS 1915+105 \cite{chou05}
in the steady `C` state.  They analyzed the steady sate data
 from the RXTE archives and found anti-correlated delay of the hard X-ray emission
 during high variability. The delays were of the order of 200 - 1000 s.  

Evidence for a truncated accretion disk was also found in the very high state
 of  XTE J1550-564 based on simultaneous ASCA - RXTE observations \cite{donkub06}.
It was found that the wide band X-ray spectrum consists of three spectral
 components (as has been found in GRS 1915+105) and they investigated three separate
 scenarios: corona encompassing the disk; truncated corona over the accretion disk but
 not coupled to it and coupled and truncated corona. From  a detailed spectral fitting
 and comparing their results to the model parameters obtained during the high-soft states,
 they concluded that the inner disk radius and/or its emission properties changed during
 the very high state, as compared to the high-soft state data.  Sriram et al. \cite{srir06}
 have analyzed this data and found anti-correlated delays in them. They also analyzed the
 power density spectra of the source which clearly shows a minute change in the
 centroid frequency of the QPO. This small change in the centroid frequency of
 QPOs indicates a physical change in the corona on a timescale of ~ 1000 s.
 The various models which explain the production of QPOs relate the origin of
 QPOs to the compact corona region close to the black hole.
 It was found that the change in the centroid frequency is strongly correlated to change
 in the soft and hard fluxes. The centroid frequency was increasing along with a
 increase in soft flux and decrease in the hard flux (see Table 1).

\begin{table}[b]
\begin{center}
\scriptsize
\begin{tabular}{|lcccc|}
\hline
\hline
Source &Delay & Flux  & QPO frequency   & kT$_{in}$ \\
 & (s)   & 10$^{-9}$ erg cm$^{-2}$ s$^{-1}$  & (Hz) & (keV)  \\
\hline
GRS 1915+105 & 704$\pm$35 & 8.50 & 3.79$\pm$0.03 & 0.83 \\
& & 6.61 & 3.41$\pm$0.02 &  \\
XTE J1550-564 & 375$\pm$13 &17.7  & 3.87$\pm$0.01 & 0.80 \\
& & 23.5 & 4.07$\pm$0.01 &  \\
& 132$\pm$9 &20.9 & 4.32$\pm$0.02 & 0.80 \\
& & 21.9 & 4.38$\pm$0.02 &  \\
\hline
\end{tabular}
\caption{Delay and QPO frequency} 
\label{microq}
\end{center}
\end{table}

\section{Compton delays.}

One of the direct methods to conclusively demonstrate the existence of a compact corona
 of the size of a few tens of Schwartzschild radius is to obtain a delay between soft
 and hard X rays due to Compton scattering. In Galactic black hole sources frequency resolved
 delays have been observed  \cite{miya88, miya91},
but the implied size is too
 large to confine a corona. Hence such low frequency delays possibly indicate the shape
 of the random shots generating the variability. The delay due to Compton scattering would
 be   a few ms in Galactic sources and a few days in quasars and a few thousand seconds in
 low mass AGNs, for sizes of a few tens of R$_S$. Detection of such delays is  observationally
 quite difficult (because of the time scales involved) and nearby bright low
 mass AGNs currently offer the best possibility.

Direct delays are observed  in Mrk 110  during a long XMM-Newton observations \cite{dasrao06}.
Mrk 110 is a nearby narrow-line Seyfert 1 galaxy and during 
the 47.4 ks of observations it showed a 10\% variation in about 3 hours.
 Cross-correlating the X-ray light curves of different energies resulted in a
 definite energy dependent delay (ranging from 200 s to 4500 s) and the delay
 was found to be proportional to the logarithm of energy (as would be expected
 if the delays are indeed due to the Comptonization process). A simple Comptonization model was
 found to be fitting the data quite well and the resultant size of the
 Compton cloud was derived to be 10 R$_S$. Spectral fitting too showed that
 the X-ray spectrum is best described by a two-component Comptonization model.
 This constitutes the best evidence for a truncated and confined accretion disk in an
 accreting black hole.

\section{Physical models}

The confinement of the Compton scattering electron cloud within a few tens of R$_S$
  is possible in  the Two Component Accretion Flow (TCAF)
  model advocated by Chakrabarti and his co-workers \cite{chak96, chaman00}.
In this model the matter accreting onto a black hole encounters a
 Centrifugal Boundary Layer (Cenbol) and radial shocks can form at the Cenbol
 (see the schematic diagram in Figure 3). The region within the Cenbol
 (region A in the figure) confines the hot Compton cloud and the
 Cenbol surface acts as a funnel (region B in the figure) for the outflow \cite{dascha99}.
In Figure 4, the 
 estimated amount of outflow as a function of Cenbol location is shown.
 This, qualitatively, explains the X-ray radio correlation as well as
 the soft and hard X-ray anti-correlation at high accretion rates. A more
 quantitative analysis, however, requires the knowledge of the
 accretion mechanism which determines viscosity or, it can also be
 parameterized as the disk and halo accretion rates - see Chakrabarti \& Titarchuk \cite{chatit95}.
A detailed wide band spectral fitting with simultaneous radio observations would
 help pin point the accretion mechanism responsible for the location of Cenbol.

The overall picture of the X-ray-radio correlation can be understood within
 the above concepts of a truncated accretion disk (at the Cenbol) whose
 radius decreases with increasing accretion disk. It is interesting to note that
  Vadawale et al. \cite{vada03} have explored this association upto the superluminal flare
 generation in GRS 1915+105.

\begin{figure}
\begin{center}
\includegraphics[angle=0,width=.8\textwidth]{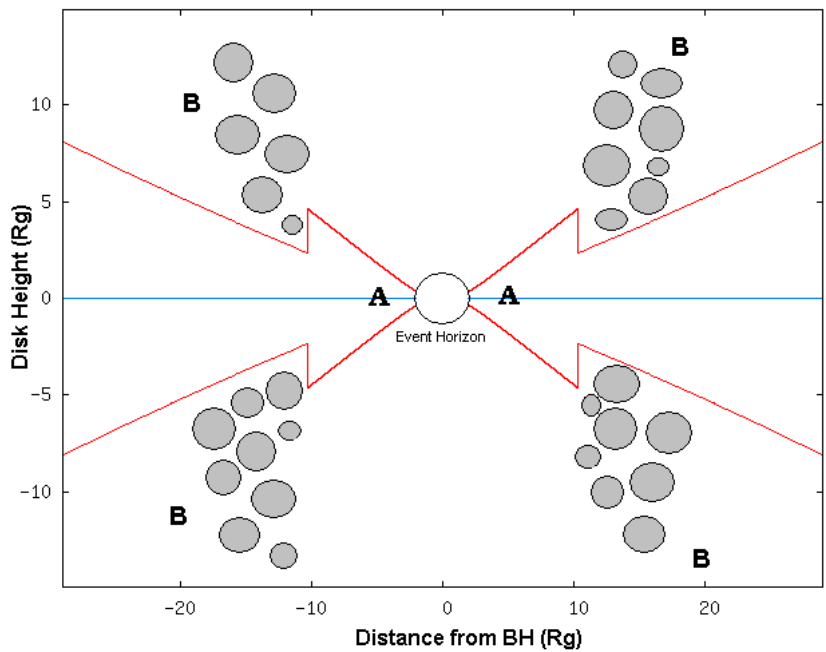}
\caption{
A schematic model invoking Cenbol to explain the X-ray radio correlation.
}
\label{cenbol}
\end{center}
\end{figure}

\begin{figure}
\begin{center}
\includegraphics[angle=270,width=.8\textwidth]{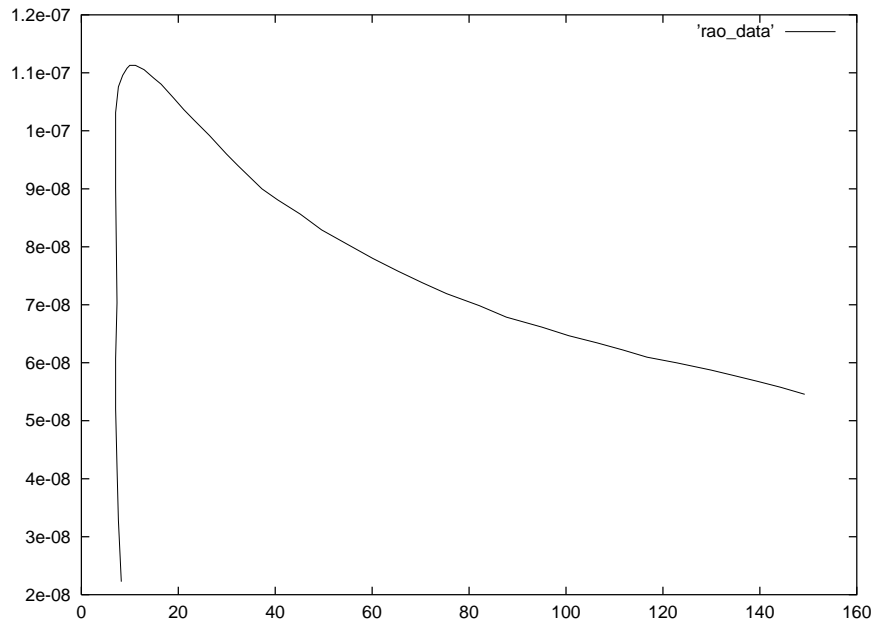}
\caption{
The outflow rate as a function of the Cenbol radius (Das 2005, private communications).
}
\label{outflow}
\end{center}
\end{figure}

\section{Future work}

The most important ingredient in understanding the disk-jet connection is getting a
 good X-ray spectroscopic measurement in  the 2 - 100 keV region and associating
 it to the radio emission. Several attempts are being made in India and these are
 briefly discussed below.

A multi-wavelength astronomy satellite called  Astrosat is being developed for a
 possible launch in the 2008-2009 time scale. It consists of three identical Large
 Area Xenon filled
Proportional Counters (LAXPC) which are similar to RXTE PCUs, but with substantially
 higher efficiency in the 10 - 50 keV region. A 1000 cm$^2$  Cadmium Zinc Telluride
 (CZT) detector array and a soft X-ray telescope (SXT) with X-ray mirrors and X-ray
  CCD operating in conjunction  with the LAXPC detectors would provide the most
 precise wide-band X-ray spectroscopic measurements. Three identical units of Sky
 Survey Monitors (SSM) which are similar in concept to the ASM onboard RXTE and UV
 imaging telescope (UVIT) will substantially enhance microquasar observing capability
 of Astrosat.

Recently, the Indian Space Research Organization  has made an Announcement of Opportunity
 to launch 100 kg class small satellites. Two proposals, made in response to this
 announcement, are particularly relevant for micro-quasar studies. The Continuous
 Spectro-Photometric Observatory for Black holes (CSPOB, PI: S. K. Chakrabarti)
 consists of Ginga type Scanning All Sky Monitors and a 400 cm$^2$ CZT detector
 for continuous tracking of select bright Black hole sources. A Super Monitor
 (SuperMon)  for X-ray Flashes, GRBs and black holes is also proposed (PI: A. R. Rao)
 which includes four  low energy proportional counters (a BATSE type XRF monitoring
 at low energies), Si-CZT combination spectroscopic detector for   identification
 and monitoring of XRFs, GRBs and black hole sources. It will have extensive onboard
 intelligence and fast satellite orientation capability and it will have the capability
 of 1 GRB/ XRF  detection every 3 days, 1 GRB/ XRF  redshift  measurement every
 fortnight, two dozen bright X-ray source tracking on semi-continuous basis,
 hundred source flux measurement on a daily basis and a dozen AGN X-ray QPO search.

\section{Conclusions}

The flux - flux correlations found between X-ray and radio emissions  indicate a
 broad trend of variations and it is quite unlikely that the  Synchrotron mechanism
 is responsible to produce both these radiations to result in a  correlation (particularly
 for bright sources). The size of a putative truncated accretion disc appears to drive
 the  X-ray-radio correlation. Precise spectral measurement in 10 - 30 keV region  and
 continuous monitoring are important to further understand this subject.  New missions
 for this purpose are proposed in India.

\section*{Acknowledgements}

It is a pleasure to thank all my collaborators who contributed to several aspects of the material presented in this article. Particular thanks are due to S.K. Chakrabarti, M. Choudhury,  T. Das, S. Dasgupta, G.C. Dewangan, S. Naik, A. Nandi, K. Sriram, J. Pendharkar, S.V. Vadawale, J. S. Yadav and A.A. Zdziarski.

\end{document}